\documentclass{aastex}
\usepackage{emulateapj5}
\usepackage{psfig,epsfig}

\shorttitle{Galaxy mapping with the SAURON spectrograph} 
\shortauthors{Roger L.\ Davies et al.}

\begin{document}

\def \eg{e.g.,\/}
\def \ie{i.e.,\/}
\def \Hb{${\rm H}\beta$}
\def \hbp{${\rm H}\beta \, ^\prime$}
\def \HgA{${\rm H}\gamma_{\rm A}$}
\def \Hg{${\rm H}\gamma$}
\def \HgF{${\rm H}\gamma_{\rm F}$}
\def \HbG{H$\beta_{\rm G}$}
\def \mgb{Mg{\,\it b}}
\def \fe{$<$Fe$>$}
\def \mgbp{Mg{\,\it b}\,$^\prime$}
\def \fep{$<$Fe\,$^\prime \negthinspace >$}
\def \kms{\rm km~s$^{-1}$}
\def \kmsM{\rm km~s$^{-1}$~Mpc$^{-1}$}
\def \prim{$^\prime$}
\def \OIIIb {[O{\small III}]$\lambda 5007$}
\def \NI {[N{\small I}]$\lambda 5199$}
\def \gon {Gonz\'{a}lez}
\def \col {CBDMSW99}
\def \tra {TFWG00}
%
%
\def\spose#1{\hbox to 0pt{#1\hss}}
\def\lta{\mathrel{\spose{\lower 3pt\hbox{$\sim$}}
    \raise 2.0pt\hbox{$<$}}}
\def\gta{\mathrel{\spose{\lower 3pt\hbox{$\sim$}}
    \raise 2.0pt\hbox{$>$}}}
%
%
\def\arcsec{\hbox{$^{\prime\prime}$}}
\def\farcm{\hbox{$.\mkern-4mu^\prime$}}
\def\farcs{\hbox{$.\!\!^{\prime\prime}$}}
\def\sauron{{\tt SAURON}} 
%
%
\title{Galaxy mapping with the SAURON integral-field spectrograph:
  \\the star formation history of NGC\,4365}
\author{Roger L.\ Davies$^1$, Harald\ Kuntschner$^1$, Eric Emsellem$^2$, R.\ Bacon$^2$,\\
  M.\ Bureau$^3$, C.\ Marcella Carollo${^4}$, Y.\ Copin$^2$,
  Bryan W.\ Miller$^{3,5}$, G.\ Monnet$^6$, \\
  Reynier F. Peletier$ ^{1,7}$, E.K.\ Verolme$^3$, P. Tim\ de Zeeuw$^3$}
\affil{$^1$ University of Durham, Department of Physics, South Road,
  Durham DH1 3LE, UK} \affil{$^2$ CRAL - Observatoire de Lyon, 9 Avenue
  Charles--Andr\'{e}, 69230 Saint-Genis-Laval, France} \affil{$^3$
  Sterrewacht Leiden, Niels Bohrweg 2, 2333 CA Leiden, The
  Netherlands} \affil{$^4$ Department of Astronomy, Columbia
  University, 538 West 120th Street, New York, NY 10027, USA}
\affil{$^5$ Gemini Observatory, Casilla 603, La Serena, Chile}
\affil{$^6$ European Southern Observatory, Karl-Schwarzschild Stra\ss e
  2, 85748 Garching, Germany} \affil{$^7$ Dept.\ of Physics \&
  Astronomy, Univ.\ of Nottingham, University Park, Nottingham NG7 2RD,
  UK} \email{Roger.Davies@durham.ac.uk}

%
%
\begin{abstract}
  We report the first wide-field mapping of the kinematics and stellar
  populations in the E3 galaxy NGC\,4365. The velocity maps extend
  previous long-slit work. They show two independent kinematic
  subsystems: the central $300 \times 700$~pc rotates about the
  projected minor axis, and the main body of the galaxy,
  $3\times4$~kpc, rotates almost at right angles to this. The
  line-strength maps show that the metallicity of the stellar
  population decreases from a central value greater than solar, to
  one-half solar at a radius of 2~kpc. The decoupled core and main body
  of the galaxy have the same luminosity-weighted age, of
  $\approx$14~Gyr, and the same elevated magnesium-to-iron ratio. The
  two kinematically distinct components have thus shared a common star
  formation history. We infer that the galaxy underwent a sequence of
  mergers associated with dissipative star formation that ended
  $\gta$12~Gyr ago. The misalignment between the photometric and
  kinematic axes of the main body is unambiguous evidence of
  triaxiality. The similarity of the stellar populations in the two
  components suggests that the observed kinematic structure has not
  changed substantially in 12~Gyr.
\end{abstract}
%
%

%
%
\section{Introduction}
\label{sec:intro}

The existence of decoupled cores in $\approx$30\% of the early-type
galaxies is strong evidence that mergers play an important part in the
evolution of these systems \citep[\eg\/][]{zf1991}. Most likely,
decoupled cores originate from the accretion of material with angular
momentum misaligned from that of the main galaxy. A few galaxies have
been studied in sufficient detail to explore when that material was
accreted or whether the event was associated with gaseous dissipation
and star formation. For example in the morphologically normal galaxies
IC\,1459 \citep{fi1988} and NGC\,5322 \citep{ben1988,rw1992} there is
no difference between the colors or line-strengths of the stellar
populations in the decoupled core and the main galaxy. In contrast, the
disturbed shell galaxy NGC\,2865 shows evidence for very recent star
formation in the decoupled component \citep*{hcb1999}.\looseness=-2

NGC\,4365 is one of the first elliptical galaxies in which minor axis
rotation was discovered \citep*{wbm1988,bsg1994}.  \citet[][hereafter
SB]{sb1995} deduced the remarkable kinematic structure of NGC\,4365
from three long-slit spectra. The main body of the galaxy rotates
around its {\em major}\/ axis, reaching a maximum velocity of
$\approx$50~\kms, whereas at smaller radii ($r=$ 2--3\arcsec) the peak
rotation velocity is 80~\kms\/ around the {\em minor}\/ axis. SB found
that the decoupled core is flatter than the main galaxy (ellipticity
$\varepsilon_{\rm core}$=0.39 cf.\ $\varepsilon_{\rm main}$=0.23). By
carrying out a careful double Gaussian analysis of the kinematics, SB
estimate that $V/\sigma= 1.3$ for the core, consistent with a disk or
bar. They deduced that the rotating core contributes 25\% of the light
within $r\approx6$\arcsec, and accounts for roughly 2\% of the {\em
  total}\/ mass. The giant elliptical galaxy in which it is embedded
has a central velocity dispersion of 275~\kms. They found that the
decoupled core contains a high-metallicity population, with an enhanced
[Mg/Fe] ratio as commonly found in giant ellipticals. SB concluded that
NGC\,4365 is triaxial, and that the formation of the core involved
substantial gaseous dissipation and star formation. However, they could
not distinguish formation scenarios where the merger(s) occur early
($z> 2$-3) from those where the mergers occur more recently
($z<1$).\looseness=-2

\begin{figure*}
\epsfig{file=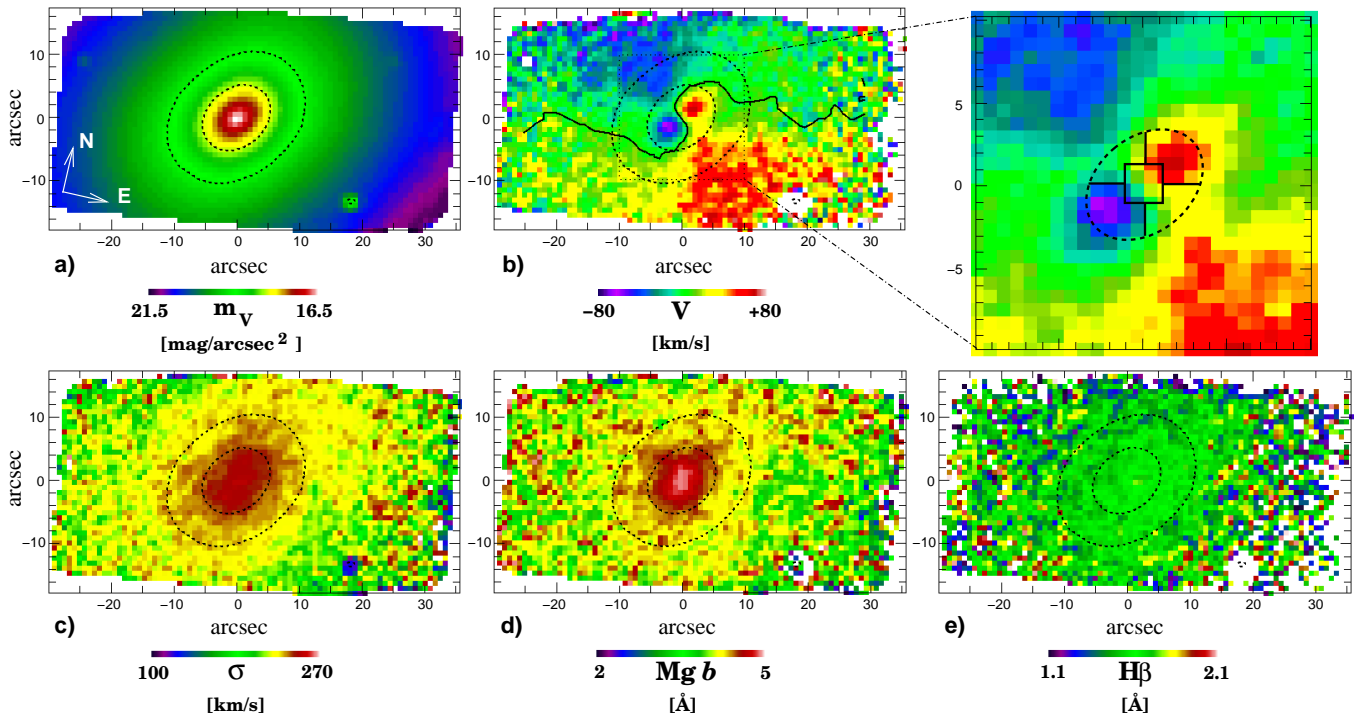}
\caption[]{\label{fig:color}Maps of a) surface brightness as
  reconstructed from our data, b) mean streaming velocity $V$, c)
  velocity dispersion $\sigma$, d) \mgb\/ line-strength, and e)
  H$\beta$ line-strength, for NGC\,4365. The maps are based on two
  partially overlapping \sauron\/ pointings of 4$\times$1800~s each,
  sampled with $0\farcs8\times0\farcs8$ pixels. Isophotal contours of
  $\mu_V = 18$ and 19 mag/arcsec$^2$ are overplotted in panels (a) to
  (e) (dashed lines). A zero-velocity contour line (solid black line)
  is shown in panel (b). The enlarged core region of the velocity map
  indicates the regions which we have used for our line-strength
  analysis: The central point is indicated by the square, the decoupled
  core by the sectors along the major axis and the main body at the
  same radius by the sectors along the minor axis.}
\end{figure*}

HST images reveal that NGC\,4365 has a smooth intensity profile with no
signs of dust, an average ellipticity of 0.26, and a very shallow
central cusp. The isophotes are disky for $1\arcsec<r<4$\arcsec\/ and
boxy at larger radii. At the very centre ($r \lta 0\farcs2$) there is a
blue point source \citep{df1995,car1997}. Deep ground-based images show
no evidence for shells or other morphological peculiarities out to
$\approx 4 \, r_e$ (Blakeslee, private communication).

Surface-brightness-fluctuation measurements indicate that NGC\,4365 is
in the Virgo W cloud beyond the main cluster \citep*{jtl1998}. We take
the distance modulus to be 31.7 mag, a distance of 22~Mpc, so that
1\arcsec\/ is $\approx$100~pc. The effective radius $r_e$ is 57\arcsec,
or $\approx$5.7~kpc \citep{bur1987}.

In this Letter, we present the first complete maps of the kinematics
and stellar populations of NGC\,4365, taken with the wide-field
integral-field spectrograph \sauron\ \citep{bac2001}. In \S2, we
briefly describe the observations. The kinematics and line-strength
index maps are presented in \S3, where we consider age-metallicity
diagnostics, the Mg-$\sigma$ diagram, and the spatial distribution of
non-solar abundance ratios. We discuss the implications of these
results for formation scenarios in \S4, and summarize our conclusions
in \S5.

\section{Observations} 
\label{sec:data}
We observed NGC\,4365 with \sauron\ mounted on the 4.2m William
Herschel Telescope on La Palma, on the nights of 29 \& 30 March 2000.
\sauron\ has a field-of-view of 33\arcsec$\times$41\arcsec, delivering
simultaneously 1431 spectra at a spectral resolution of 3.6~\AA\/
(FWHM), $0\farcs95\times0\farcs95$ spatial sampling, and 100\% sky
coverage \citep{bac2001}. Another 146 spectra are taken $1\farcm9$ away
from the main field, to allow accurate sky subtraction. The wavelength
range of the current setup is $\approx$4810--5350 \AA. We observed two
fields overlapping by $\approx$20\arcsec\/ on the nuclear region of
NGC\,4365, each field having four separate exposures of 1800~s
(dithered by $\approx$1\arcsec, \ie\/ one lenslet). The combined
datacubes cover a total region of 33\arcsec$\times$ 63\arcsec\/ on the
sky. The small offsets between the four 1800~s integrations at each
position enable us to re-sample the final datacubes onto
$0\farcs8\times0\farcs8$ pixels (drizzling technique). The seeing of
the merged datacube was measured on three point-like objects in the
reconstructed image. It is homogeneous over the field with a value of
$1\farcs6 \pm 0\farcs1$ (FWHM).  We took arc-lamp spectra before and
after each individual exposure for accurate wavelength calibration.

We reduced the raw \sauron\ exposures by means of the algorithms
described in \citet{bac2001}. We used the individually-extracted,
wavelength-calibrated and continuum-corrected spectra to derive the
stellar kinematics and line-strength indices as a function of
(two-dimensional) position in NGC\,4365. We measured the mean velocity
$V$ and the velocity dispersion $\sigma$ with the Fourier Correlation
Quotient method \citep{ben1990}, and obtained the line-strength indices
\Hb, \mgb\/ and Fe5270 in the Lick/IDS system \citep{wor1994}, taking
into account the internal velocity broadening and differences in the
instrumental resolution.

\begin{figure*}
  \epsfig{file=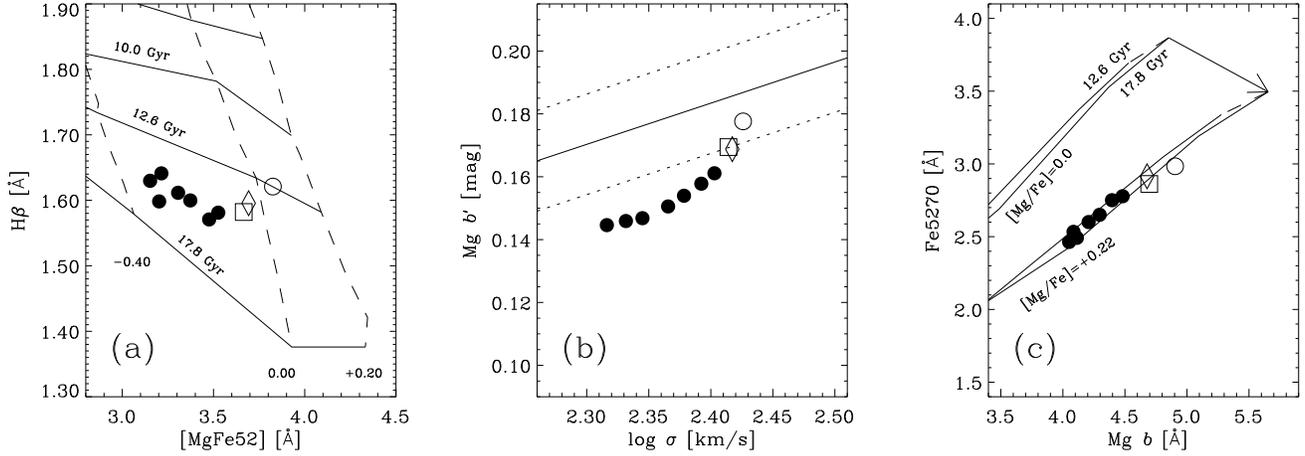}  
  \caption[]{\label{fig:age_sig}(a) [MgFe52] {\em vs}\/ H$\beta$
    equivalent width diagram. The open circle represents the average
    line-strength of the very central data points ($r < $1\farcs6), the
    open diamond represents the region of the decoupled core, and the
    open square reflects the mean of the data in the main body of the
    galaxy at the same radii as the decoupled core but along the minor
    axis (see also Figure~1, enlarged core region). For larger radii,
    the data was averaged in elliptical annuli centered on the
    photometric nucleus (filled circles); mean semi-major axis radii:
    5.6, 7.2, 9.6, 12.9, 16.0, 19.9 \& 26.1\arcsec\/ (radius increases
    from right to left).  Overplotted are the predictions of stellar
    population models from \citet{vaz1999}. The solid lines are lines
    of constant age and the dashed lines are lines of constant
    metallicity ([M/H]). (b) Local velocity dispersion {\em vs}\/ local
    \mgbp\/ index. The solid line indicates the average relation for
    the cores of early-type galaxies (Colless et~al.  1999) and the
    dashed lines indicate the 1$\sigma$ spread around it. (c) \mgb\/
    {\em vs}\/ Fe5270 equivalent width diagram. Overplotted are stellar
    population models from \citet{vaz1999} with solar abundance ratios
    for an age of 12.6 \& 17.8 Gyrs. Additionally a correction of the
    models for {\em non-solar}\/ abundance ratios of
    [Mg/Fe]~$=+0.22$~dex is shown (Trager et~al. 2000). The error bars
    on the mean linestrength in a given zone are omitted for clarity as
    they are smaller or similar to the size of the symbols in the
    diagrams.}
\end{figure*}

\section{Results}
Figure~\ref{fig:color}{\em a}\/ shows the surface brightness
distribution of NGC\,4365, as reconstructed from our
spectra.\notetoeditor{The color Figure should appear on the top (2
  column width) of the second page} The reconstructed intensity map
agrees well with the HST image after seeing convolution and binning to
the \sauron\/ spatial sampling.

Figure~\ref{fig:color}{\em b}\/ shows the spectacular kinematically
decoupled core of NGC\,4365 in detail. The core extends
$\approx$7\arcsec$\times$3\arcsec\/ and rotates about the minor axis.
The maximum observed core rotation speed is 80 \kms\/ at a radius of
2\farcs2. The main body of the galaxy rotates slowly about an axis
misaligned by $8\pm2^{\circ}$ with the major axis. The rotation
velocity rises to 45 \kms\/ at $r = 7$\arcsec, and remains constant at
larger radii. The velocity field of the main body is not symmetric
about the minor axis and the loci of zero velocity (shown as bold line
in Figure 1b) and maximum velocity, are not perpendicular. We will
explore the consequences of this in a later paper presenting dynamical
models. The velocity dispersion falls off smoothly from its central
maximum of 275~\kms, and the contours of constant dispersion follow the
isophotes (Figure~\ref{fig:color}{\em c}). A detailed comparison shows
excellent agreement with the SB long-slit data \citep{zee2001}.

Figure~\ref{fig:color}{\em d}\/ shows that the distribution of \mgb\/
has a central peak, whereas the \Hb\/ value is roughly constant across
the galaxy (Figure~\ref{fig:color}{\em e}).  Our average value for the
central \Hb\/ absorption strength is $1.61\pm0.04$~\AA\/, in good
agreement with the Lick/IDS measurement of $1.66\pm0.21$~\AA\/
\citep{tra1998}. Furthermore, we find no indication for either \OIIIb\/
or \Hb-emission in our spectra, so there is no evidence that our age
estimates are affected by nebular emission.

In Figure~\ref{fig:age_sig}{\em a}\/ we show the [MgFe52] {\em vs}\/
\Hb\/ age-metallicity diagnostic diagram ([MgFe52]~=~$\sqrt{{\rm Mg}\,
  b \times {\rm Fe5270}}$)\notetoeditor{Figure 2 should appear on the
  top (2 column width) of the third page}. In order to probe the
stellar populations of the decoupled core with respect to the main
galaxy we have averaged the line-strength in certain key regions (see
also Figure~\ref{fig:color}, enlarged core region). The very central
region ($r<$1\farcs6) of the galaxy is represented by an open circle.
Furthermore, we have identified the decoupled core using the velocity
maps and plot the average value for the line-strengths in this region
as an open diamond.  Comparing this with the average line-strengths of
the main galaxy at the {\em same}\/ radius along the minor axis (open
square) we find that these two kinematically distinct regions have
identical stellar populations. At larger radii ($r > $ 5\farcs0) we
averaged all lenslets in elliptical annuli (filled circles). The metal
line-strength decreases with increasing radius and there is a small
increase in \Hb\/ absorption strength.

In order to make age and metallicity estimates, we use the
\citet{vaz1999} models, which utilize the empirical stellar library of
\citet{jon1997} to predict line-strengths for a single-burst stellar
population as a function of age and metallicity. The models were
smoothed to the Lick/IDS resolution and include improved stellar
population parameters (Vazdekis 2001, in preparation). The model
predictions are shown in Figure~\ref{fig:age_sig}{\em a}\/ \& {\em c}.
The central metallicity is estimated to be 1.15~$Z_{\odot}$, decreasing
towards larger radii ($\approx$0.3 dex per dex in radius) at a roughly
constant age of 14~Gyr (Figure~\ref{fig:age_sig}{\em a}). We note that
the absolute age calibration of the models remains subject to systmatic
errors, but all our conclusions are based on relative age differences
which are much more robust. The small increase in H$\beta$ absorption
at the very center ($r\lta 1\farcs6$) suggests a luminosity-weighted
age of $\approx$12~Gyr. We can account for this by superimposing a
younger population on that of the main body: 6\% of the mass in a
stellar population with the same metallicity and an age of 5 Gyrs is
sufficient.




Figure~\ref{fig:age_sig}{\em b}\/ shows the \mgb\,--$\sigma$ relation
within NGC\,4365. The central data points agree well with the relation
for the cores of early-type galaxies \citep{col1999}, suggesting that
the core properties of NGC\,4365 are similar to those of other
ellipticals.  For $r \lta 6$\arcsec, the local \mgb--$\sigma$ relation
shows a steeper slope than the global relation, but overall the
gradient in this diagram is typical of similar galaxies studied by
\citet*{dsp1993} and \citet*{cd1994}.

In Figure~\ref{fig:age_sig}{\em c}\/ we plot \mgb\/ {\em vs}\/ Fe5270.
Stellar population models \citep{vaz1999} at solar abundance ratios and
for ages 12.6 and 17.8 Gyr are overplotted. In these coordinates the
effects of age and metallicity are almost completely degenerate hence
the model predictions overlap. Consistent with other giant ellipticals
\citep[see \eg][]{kun1998,kun2001}, the data points for NGC\,4365 lie
off the solar ratio models towards larger values of \mgb\/ and lower
Fe5270 line-strength. Using the corrections given by \citet{tra2000} we
also plot stellar population models at [Mg/Fe]~=~0.22~dex, which are a
good representation of the whole of NGC\,4365. There is no difference
between the decoupled core region and the main body of the galaxy.  SB
find that the magnesium-to-iron ratio is further enhanced in the very
center.  Our data, whilst marginally consistent with theirs, indicate
no additional enhancement.

\section{Discussion}
We now explore how the {\tt SAURON} two-dimensional line-strength maps
constrain the star formation history of both the main body and the core
of NGC\,4365. There is a dramatic difference in the kinematics of the
two regions of the galaxy, but other properties suggest that NGC\,4365
is a normal elliptical galaxy and that the core and main body had a
common star formation history. Furthermore, the K-band surface
brightness fluctuations in NGC\,4365 place it amongst the old metal-rich
ellipticals \citep{jtl1998}.

The elevated magnesium-to-iron ratio is roughly constant across the
entire galaxy (a region of 4$\times$3~kpc). Such non-solar abundance
ratios arise in populations enriched primarily with the products of
Type II supernovae, either in a rapid initial burst of star formation
or one skewed to massive stars \citep{wor1998}. The uniformity of the
elevated magnesium-to-iron ratio also suggests that the whole galaxy
experienced a common star formation history, involving considerable
gaseous dissipation, thus generating the high central metallicity and
inward metallicity gradient. A possible formation scenario would be the
merger of gas-rich fragments at high redshift.  Such an event would be
modest compared to the rates of star formation inferred for
high-redshift sub-millimeter galaxies \citep[\eg][]{ivi2000}.  The
decoupled core could originate from stars ejected into a tidal tail
(with the appropriate angular momentum) as a result of a major merger
that formed the bulk of the stars. These stars fall back to produce the
kinematically distinct component at the centre.

If star formation was taken to completion and the residual gas
exhausted roughly 12~Gyr ago, then the decoupled kinematic structure in
NGC\,4365 must be long-lived. The misalignment of the kinematic and
photometric axes show that the main body of the galaxy is triaxial,
with the bulk of its stars on long-axis tubes and the stars in the core
predominantly on short-axis tubes \citep{sta1991,azh1994}. This is
similar to the structure inferred for, \eg\/ NGC\,4261 \citep{db1986}
and NGC\,4406 \citep*{fih1989}. The full two-dimensional structure and
kinematics derived from the \sauron\ data, when combined with our
dynamical models \citep[see \eg\/][]{cre1999}, will enable us to
distinguish between a thin disk or a thick structure for the core.

\section{Conclusions}

The \sauron\ maps present a complete view of the kinematics and stellar
populations of NGC\,4365. They show two independent kinematic
subsystems: the central $300 \times 700$~pc and the main body of the
galaxy, rotating almost at right angles to each other. The misalignment
of $82^{\circ}$ between the photometric and kinematic axes of the main
body is unambigous evidence of triaxiality. The \sauron\ maps enable us
to compare the stellar population in the decoupled component with that
in the main body of the galaxy at the same radius. We find these
populations to be indistinguishable in age, metallicity and abundance
ratios. We find an age of $\approx$14~Gyr and a decrease in
metallicity, from larger than solar in the center to half solar at a
radius of 2~kpc ($\approx0.4 \, r{_e}$). We suggest that NGC\,4365
underwent dissipative star formation at high redshift, probably through
one or more mergers.  Later generations of stars formed a more
centrally-concentrated, metal-enriched stellar population. Star
formation was complete and the residual gas was exhausted roughly
12~Gyr ago. This also suggests that the observed kinematics and
triaxial structure is stable.

\acknowledgements It is a pleasure to thank Rene Rutten and the ING
staff, in particular Tom Gregory, for support on La Palma. We thank
Richard McDermid for assistance during the observing run. The \sauron\/
project is made possible through grants 614.13.003 and 781.74.203 from
ASTRON/NWO, PPARC grant ``Extragalactic Astronomy \& Cosmology at
Durham 1998-2002'' and financial contributions from the Institut
National des Sciences de l'Univers, the Universit\'e Claude Bernard
Lyon I, together with the universities of Durham and Leiden.  RLD
gratefully acknowledges the award of a Research Fellowship from the
Leverhulme Trust.

\clearpage
\end{document}